\documentclass[conference]{IEEEtran}
\IEEEoverridecommandlockouts
\ifCLASSINFOpdf
\usepackage[pdftex]{graphicx}
\else
\usepackage[dvips]{graphicx}
\fi
\usepackage{cite}
\usepackage{amsmath,amssymb,amsfonts}
\usepackage{graphicx}
\usepackage{mathtools}
\usepackage{textcomp}
\usepackage{amsfonts} 
\usepackage{algpseudocode} 
\usepackage{amsmath}
\usepackage[ruled,vlined]{algorithm2e}
\usepackage{booktabs}
\usepackage{xcolor}
\def\BibTeX{{\rm B\kern-.05em{\sc i\kern-.025em b}\kern-.08em
    T\kern-.1667em\lower.7ex\hbox{E}\kern-.125emX}}

\begin{document}
			\title{Real-Time Energy Management Strategies for Community Microgrids
}

%

\author{Moslem~Uddin, 
	Huadong Mo, 
	Daoyi Dong

	\thanks{Moslem Uddin is with 
		School of Engineering \& Technology, The University of New South Wales, Canberra, ACT 2610, Australia (email: moslem.uddin.bd@gmail.com).}
	\thanks{Huadong Mo is with School of Systems and Computing, The University of New South Wales, Canberra, ACT 2610, Australia (email: huadong.mo@adfa.edu.au).}
	\thanks{Australian AI Institute, FEIT, University of Technology	Sydney, Sydney, NSW 2007, Australia Australia (email: daoyidong@gmail.com).}
}

\maketitle

\begin{abstract}
This study presents a real-time energy management framework for hybrid community microgrids integrating photovoltaic, wind, battery energy storage systems, diesel generators, and grid interconnection. The proposed approach formulates the dispatch problem as a multi-objective optimization task that aims to minimize operational costs. Two control strategies are proposed and evaluated: a conventional rule-based control (RBC) method and an advanced deep reinforcement learning (DRL) approach utilizing proximal policy optimization (PPO). A realistic case study based on Australian load and generation profiles is used to validate the framework.
Simulation results demonstrate that DRL-PPO reduces operational costs by 18\%,  CO\textsubscript{2} emissions by 20\%, and improves system reliability by 87.5\% compared to RBC. Beside, DRL-PPO increases renewable energy utilization by 13\%, effectively reducing dependence on diesel generation and grid imports.
These findings demonstrate the potential of DRL-based approaches to enable cost-effective and resilient microgrid operations, particularly in regional and remote communities.
\end{abstract}


\section{Introduction}
The global transition toward low-carbon and decentralized energy systems has significantly elevated the role of microgrids (MGs) as a core component of resilient and sustainable energy infrastructure \cite{Schutz2018,Laszka2017,Lan2019}.
These systems are particularly advantageous in areas where the grid connections are unreliable, intermittent, or absent \cite{Zubieta2016,Guo2018,Zhang2018}. 
MGs operate by integrating various distributed energy resources (DERs), including solar PV systems, wind turbines (WTs), diesel generators (DGs), and battery storage units. These components collectively meet local electricity demands while enhancing system reliability. Furthermore, MGs contribute to carbon emission reduction and promote energy independence \cite{uddin2025cost,Eyimaya2024,Chaudhary2021,Asano2024}. 

The effective operation of MGs relies heavily on energy management systems (EMSs) that can optimize real-time dispatch of energy resources. This involves responding to dynamic load profiles, variability in renewable generation, and fluctuations in grid interaction costs \cite{Moazzen2024,Ma2024,Neelashetty2025}.
Traditional energy management strategies have historically relied on heuristic methods, including RBC systems, fuzzy-logic algorithms, and programmable logic controllers. These methodologies have been widely adopted due to their ease of implementation, low computational demands, and operational simplicity \cite{Bakare2024, Yu2016, Ma2018, Pinthurat2023, Zhou2018}. Based on expert knowledge and empirical observations, these systems demonstrate a reliable performance under stable and predictable operating conditions \cite{uddin2025integrated}. However, these conventional approaches encounter significant challenges  under dynamic and uncertain operating conditions. The effectiveness of rule-based systems is significantly diminished under scenarios with high renewable energy penetration or highly variable load demand. In such cases, their rigid structure limits adaptability to dynamic and uncertain operating conditions \cite{Liu2018, Shahzad2022, Aatabe2024, Niknami2024, Moosavi2022}.

Recent advancements have focused on DRL frameworks to enhance predictive control and energy management in MGs. 
A two-step diffusion policy DRL method was introduced to manage hybride MGs, incorporating carbon emission trading and green certificate trading to guide energy behaviors and reduce emissions \cite{Zhang2024Two-Step}. Pan et al. \cite{Pan2024Online} employed a DRL-based soft actor–critic (SAC) algorithm to optimize EV battery utilization in vehicle-to-grid scheduling, enhancing profitability and sustainability in multi-energy MG operations. Furthermore, a DRL-based SAC method was proposed to optimize MG operation costs by leveraging load flexibility, demonstrating significant reductions in grid dependence and operational expenses \cite{Pei2024Deep}.
Despite certain limitations, these studies provide valuable insights into the application of MGs and EMSs in various geographical and contextual settings, which can be applied to the design and implementation of advanced EMSs for MGs in remote areas.
\begin{figure*}[!tp]
	\centering
	\includegraphics[width= 5.8 in]{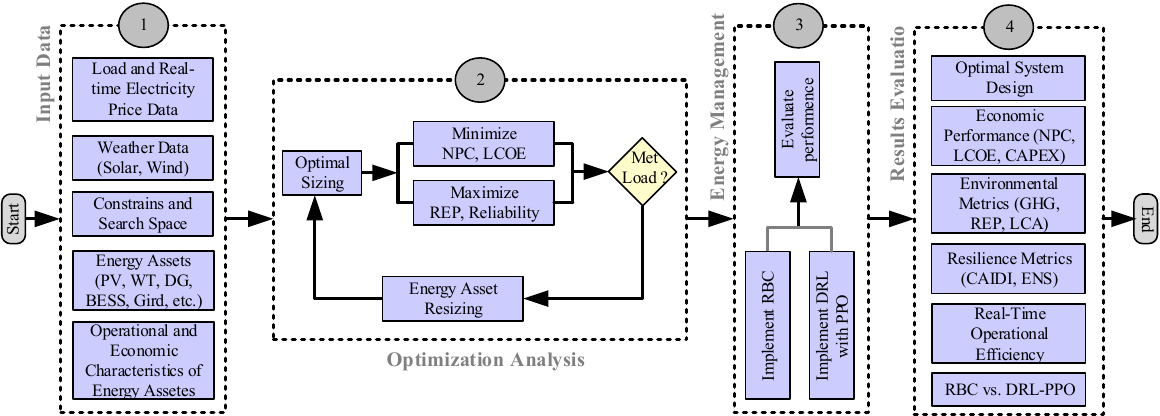}
	\caption{Proposed Methodology for Real-Time Energy Management in  Community MGs.}
	\label{fig_flow}
\end{figure*}

Although DRL-based EMSs have  demonstrated potential in offline or planning-stage MG optimization, their real-time applicability under practical scenarios, particularly in Australian contexts with hybrid energy configurations, remains underexplored. Therefore, this study investigates and compares the performance of RBC and DRL-PPO for the real-time energy management of a hybrid community MG. The energy management problem is formulated as a multi-objective optimization task. The objective is to minimize the total operational cost, which comprises grid interaction costs, battery degradation costs, and diesel fuel costs.
A realistic case study based on time-series data from Central Tilba, New South Wales, is employed to validate the proposed framework. This evaluation demonstrates its practical relevance in the Australian context.
In summary, the main contributions and originality of this study are as follows: 

%
%
\begin{itemize}
	\item 
	A unified simulation framework is developed for benchmarking heuristic (RBC) and intelligent (DRL-PPO) energy management strategies under realistic operating conditions and constraints.
	\item 
	An adaptive DRL-PPO agent is trained and deployed for real-time dispatch, with the ability to dynamically optimize control decisions based on system states such as battery state-of-charge (SOC), renewable availability, demand, and grid status.
	\item A detailed comparative analysis is conducted across five key performance indicators (KPIs): system reliability, operational cost, renewable energy utilization, battery cycling, and energy self-sufficiency.
	\item A case study of an Australian hybrid MG demonstrates the real-world applicability of the proposed DRL-based EMS and its potential to support sustainable energy transitions in remote and regional communities.
\end{itemize}

The remainder of this chapter is organized as follows. The methodology employed in this study is detailed in Section II, and the results are presented in Section III. Finally, Section IV presents the findings and conclusions drawn from the investigation.

\section{Methods}\label{Sec II}
This section outlines the methodological framework for the real-time energy management of a hybrid community MG, integrating optimal system sizing with rule-based and intelligent control strategies. The proposed framework is illustrated in Fig.  \ref{fig_flow}.
This study focuses on real-time energy dispatch for a grid connected community MG. The system configuration is initially derived using HOMER Pro for optimal sizing, considering the technical, economic, and environmental objectives. The resulting Mg architecture integrates multiple DERs and storage systems to ensure high reliability, cost-effectiveness, and carbon reduction.
In the subsequent stage, the optimized MG configuration is integrated with an energy management model to enhance operational performance. 
Two distinct energy management strategies are implemented and compared: a traditional RBC system and an advanced DRL-PPO model. The RBC method relies on predetermined static protocols for managing battery usage, interacting with the power grid, and activating DGs. By contrast, the DRL-PPO technique creates flexible control mechanisms that adjust in real time to changes in electricity costs, energy consumption patterns, and the production of RESs.
The performances of the RBC and DRL-PPO are evaluated based on real-time operational efficiency, including battery utilization, renewable utilization, and resilience metrics. 

\subsection{System Configuration and Modeling}
The hybrid MG considered in this study includes PV arrays, WTs, DGs, battery energy storage systems, and grid interconnection. The system configuration is initially optimized using HOMER Pro software to ensure techno-economic feasibility under given environmental and load conditions. 
A mathematical approach is used to determine the system components in terms of the generation and expenses. 

The PV array is modeled in such a manner that the components generate power directly proportional to the normalized solar irradiance, capacity of the PV array, and its derating factor. 
However, it is assumed that the impact of temperature on the PV output is minimal. 
The following equation is used to determine the output of a PV module at a specific time $ t $, with a rated capacity of $P^r_{pv}$, a derating factor of $\beta$, and normalized irradiance of solar $G_t$ at time $ t $ \cite{uddin2024energy}:
\begin{equation}
	P_{\text{pv}}(t) = P^r_{\text{pv}} \times G_t \times \beta; \ \ \ \ \  \, G_t = \frac{G}{G_0}
\end{equation}
where $ G $ refers to the actual solar radiation incident on the PV array (in $ kW/m^2 $) at a given moment, and  $ G_0 $ denotes the incident radiation at standard test conditions, typically defined as 1 $ kW/m^2 $.

The calculation of the maximum power output ($ P_w $) of a WT is influenced by several variables, including air density ($\rho$), turbine swept area (A), power coefficient ($\mu$), and wind velocity (v). This can be expressed as follows \cite{hassan2023review}:
\begin{equation}
	P_w = 0.5 \times \rho \times A \times \mu \times V^3.
\end{equation}

As a backup power source, DG is designed to operate according to its fuel consumption pattern, which is directly proportional to the electrical output, as indicated by Eq. (\ref{Eq:DG}). The hourly fuel consumption of the generator can be mathematically represented by a linear relationship \cite{uddin2023microgrids}. The model is based on the power requirements of the load. 
Therefore, this correlation can be articulated as:

\begin{equation}\label{Eq:DG}
	k_{c,fuel}(t) = \eta_a \times P^o_{dg} + \eta_b \times P^r_{dg} 
\end{equation}
where $ \eta_a $ refers to the generator fuel curve slope and $ \eta_b $ denotes the generator fuel intercept coefficient, which are provided by the manufacturer. In addition, $ P^o_{dg} $ represents the generated power output of the DG in kW at a specific time step (t), whereas $ P^r_{dg} $ is the rated capacity of the DG in kW. 

The stored energy in the battery is determined by several factors, including the depth of discharge (DoD) and overall capacity of the battery ($ C_{bat} $). This relationship can be expressed as follows:
\begin{equation}
	E_{stored}=DoD \times C_{bat}.	
\end{equation}

Selecting an appropriate power converter for an MG requires ensuring its capacity to handle the maximum anticipated AC load. This load can be determined using Eq. (\ref{Eq:con}), which calculates the output power based on the input power ($ P_{in} $) and inverter efficiency ($ \eta $), considering the maximum load demand and inverter specified efficiency.
\begin{equation} \label{Eq:con}
	P_{out}= \eta \times P_{in}.
\end{equation}

\subsection{Energy Management Problem Formulation}
The real-time dispatch problem is modeled as a finite-horizon Markov Decision Process (MDP),  defined by the tuple $\langle \mathcal{S}, \mathcal{A}, \mathcal{P}, \mathcal{R}, \gamma \rangle$. 
In this framework, the agent interacts with the environment to select optimal actions that minimize the expected cumulative operational cost over a finite time horizon $T$.  
The total operational cost includes grid interaction costs ($C_{grid}$), battery degradation costs ($C_{deg}$), and diesel fuel costs( $C_{dg}$). The objective function is defined as:
	
\begin{equation}
	\min_{\pi} \ \mathbb{E}_{\pi} \left[ \sum_{t=1}^{T} \left( C_{grid}(t) + C_{deg}(t) + C_{dg}(t) \right) \right]
\end{equation}
	\[
\begin{aligned}
	\text{s.t.} \quad 
	& 0 \leq SOC(t) \leq SOC_{\max}, \quad \forall t \in [1, T]  \\
	& P_{\text{load}}(t) = P_{\text{pv}}(t) + P_{\text{w}}(t) + P_{\text{bat}}(t) + P_{\text{dg}}(t) + P_{\text{grid}}(t)\\
	& P_{\text{bat}}(t) = P_{\text{bat}}^{\text{dis}}(t) - P_{\text{bat}}^{\text{ch}}(t) \\
	& P_{\text{grid}}(t) = P_{\text{grid}}^{+}(t) - P_{\text{grid}}^{-}(t) \\
	& \zeta(t) \sim \text{Bernoulli}(p_{\text{out}}), \quad \text{with } p_{\text{out}} = 0.01\\
	& P_{\text{grid}}^{+}(t) \cdot \zeta(t) = 0 \\
	& 0 \leq P_{\text{pv}}(t) \leq P_{\text{pv}}^{\max}(t) \\
	& 0 \leq P_{\text{w}}(t) \leq P_{\text{w}}^{\max}(t) \\
	& 0 \leq P_{\text{dg}}(t) \leq P_{\text{dg}}^{\max} \\
	& 0 \leq P_{\text{bat}}^{\text{ch}}(t) \leq P_{\text{bat}}^{\max} \\
	& 0 \leq P_{\text{bat}}^{\text{dis}}(t) \leq P_{\text{bat}}^{\max}
\end{aligned}
\]
where $\pi: \mathcal{S} \rightarrow \mathcal{A}$ is the policy mapping observed states to actions.  Meanwhile, grid outages are modeled as a Bernoulli process with a 1\% probability of occurrence at each time step.

\subsection{RBC-based Dispatch Strategy}
The RBC framework utilizes a set of predefined heuristic rules to coordinate the operation of DERs. The strategy prioritizes renewable generation, followed by BESS and DG, depending on availability and system constraints. The control logic is implemented to support peak shaving, load shifting, and energy arbitrage. Three operational scenarios are considered: grid-connected normal operation, partial RES support with battery backup, and islanded operation with DG dispatch. The full algorithmic flow is presented in Algorithm 1.
The BES is utilized for energy arbitrage, peak shaving, and smoothing of intermittent renewable generation. When the renewable generation exceeds the load demand, excess energy is utilized to charge the battery, and any remaining surplus is exported to the grid. Conversely, during periods of insufficient renewable generation, the battery discharges to compensate for the load demand. Any residual deficit is subsequently met through grid imports or, if necessary, by dispatching the diesel generator as a backup source. 

\begin{algorithm}[h]
	\caption{RBC-Based Energy Dispatch}
	\label{Alg_RBC}
	\small
	\SetKwInOut{Input}{Input}
	\SetKwInOut{Output}{Output}
	
	\Input{
		Time step $t$, load demand $P_{\text{load}}(t)$, PV generation $P_{\text{pv}}(t)$, wind generation $P_{\text{w}}(t)$, battery limits $P_{\text{bat}}^{\max}$, SOC limit $SOC(t)$, grid status $G_{\text{flag}}(t)$, grid capacity $P_{\text{grid}}^{\max}$, DG capacity $P_{\text{dg}}^{\max}$
	}
	\Output{
		Battery action, grid import/export, DG operation, SOC, cost metrics
	}
	
	\For{each time step $t$}{
		\uIf{$G_{\text{flag}}(t) = 1$ (grid available)}{
			\uIf{$P_{\text{load}}(t) < P_{\text{pv}}(t) + P_{\text{w}}(t)$}{
				Compute surplus: $P_{\text{sur}} = P_{\text{pv}} + P_{\text{w}} - P_{\text{load}}$ \\
				\uIf{$P_{\text{sur}} \leq P_{\text{bat}}^{\max,\text{ch}}$}{
					Charge battery with $P_{\text{sur}}$
				}
				\Else{
					Charge battery with $P_{\text{bat}}^{\max,\text{ch}}$ \\
					Export remaining power to grid
				}
				Record grid export
			}
			\Else{
				Compute deficit: $P_{\text{def}} = P_{\text{load}} - (P_{\text{pv}} + P_{\text{w}})$ \\
				\uIf{$P_{\text{def}} \leq P_{\text{bat}}^{\max,\text{dis}}$ and $SOC(t) > 0$}{
					Discharge battery
				}
				\Else{
					Import remaining deficit from grid (up to $P_{\text{grid}}^{\max}$)
				}
				Record SOC and grid import
			}
		}
		\Else{ 
			\uIf{$P_{\text{load}} > P_{\text{pv}} + P_{\text{w}} + P_{\text{bat}}^{\max,\text{dis}}$}{
				Use DG to meet remaining demand (up to $P_{\text{dg}}^{\max}$)
			}
			\Else{
				Discharge battery to meet load
			}
			Record DG usage and cost
		}
	}
	Record overall performance metrics
\end{algorithm}

\begin{algorithm}[h]
	\caption{Proposed DRL-PPO Framework for MG Energy Management.}
	\label{Alg_PPO}
	\small
	\SetKwInOut{Input}{Input}
	\SetKwInOut{Output}{Output}
	
	\Input{MG environment with system parameters and real-time data}
	\Output{Trained PPO policy $\pi_\theta^*$ for real-time energy management}
	
	\textbf{Initialize} PPO agent with policy network $\pi_\theta$ \\
	
	\For{each episode $e = 1, 2, \dots, E$}{
		Reset environment: $s_0 \gets \texttt{env.reset()}$ \\
		\For{each time step $t = 1, 2, \dots, T$}{
			Observe state: $s_t = [SOC_t, h_t, P_{pv}(t), P_w(t), P_{load}(t), G_{\text{flag}}(t)]$ \\
			Select action: $a_t = [P_{\text{bat}}(t), P_{\text{dg}}(t)] \sim \pi_\theta(s_t)$ \\
			Clip actions:
			\[
			P_{\text{bat}}(t) \in [-P_{\text{bat}}^{\max,\text{ch}}, P_{\text{bat}}^{\max,\text{dis}}],
			\quad
			P_{\text{dg}}(t) \in [0, P_{\text{dg}}^{\max}]
			\]
			Compute net power balance:
			\[
			P_{\text{net}}(t) = P_{pv}(t) + P_w(t) + P_{\text{bat}}(t) + P_{\text{dg}}(t) - P_{load}(t)
			\]
			Compute grid interaction cost:
			\[
			C_{\text{grid}}(t) =
			\begin{cases}
				- P_{\text{net}}(t) \cdot C_{\text{sell}}, & \text{if } P_{\text{net}}(t) > 0 \\
				P_{\text{net}}(t) \cdot C_{\text{grid}}, & \text{if } P_{\text{net}}(t) < 0
			\end{cases}
			\]
			Compute cost function and reward:
			\[
			r_t = - \left( C_{\text{grid}}(t) + C_{\text{dg}}(t) + C_{\text{deg}}(t) \right)
			\]
			Store transition $(s_t, a_t, r_t, s_{t+1})$ in buffer \\
			\If{episode ends or batch size threshold reached}{
				Estimate advantage using Generalized Advantage Estimation (GAE) \\
				Update policy $\pi_\theta$ via PPO clipped surrogate objective
			}
			Update state: $s_t \gets s_{t+1}$
		}
	}
	
	\textbf{End Training Phase} \\
	
	\textbf{Evaluation Phase:} \\
	Reset environment: $s_0 \gets \texttt{env.reset()}$ \\
	\For{each evaluation step $t = 1, 2, \dots, T$}{
		Observe state $s_t$ \\
		Select action: $a_t = \pi^*_\theta(s_t)$ using trained policy \\
		Execute action and record performance metrics (e.g., cost, reliability, renewable utilization)
	}
\end{algorithm}

\subsection{Proposed DRL Framework}
To enhance the adaptability of MG EM, this study integrates PPO, a policy-gradient reinforcement learning algorithm, into the dispatch strategy. Unlike rule-based methods, PPO learns an optimal control policy through interactions with the environment, optimizing energy storage utilization, grid imports/exports, and DG usage dynamically. The proposed DRL-PPO energy management algorithm is designed to optimize and regulate the energy flows within a community MG incorporating diverse energy sources, including RESs, battery storage, and conventional power sources.  The details of the implemented energy-dispatch strategy are presented in Algorithm \ref{Alg_PPO}.

\subsubsection{State Space}
The state vector at time $t$ is defined as:
\[
s_t = [SOC(t), h(t), P_{pv}(t), P_{w}(t), P_{load}(t), G_{flag}(t)]
\]

\subsubsection{Action Space}
The action vector at time $t$ is defined as:
\[
a_t = [\Delta P_{bat}, P_{dg}]
\]

\subsubsection{Reward Function}
The reward function is designed to minimize operational costs:
\[
r_t = -(C_{grid} + C_{deg} + C_{dg})
\]

\subsubsection{Training Methodology} 
The PPO agent is trained in a simulated environment using real-world MG data. Subsequently, the learned policy is deployed to optimize the real-time dispatch. The PPO training process comprises the following steps.

\begin{itemize}
	\item Collecting trajectories using the current policy.
	\item Computing advantage estimates using GAE.
	\item Updating the policy using stochastic gradient descent.
	\item Iterating until convergence.
\end{itemize}

\subsubsection{PPO Algorithm}

The PPO is selected owing to its stable policy updates, robustness in continuous action spaces, and sample efficiency. The optimization problem is addressed using the PPO clipped objective function, as defined in Eq. \ref{PPO_clipped}.
\begin{equation}\label{PPO_clipped}
	L(\theta) = \mathbb{E}_t \left[ \min \left( r_t(\theta) A_t, \text{clip} \left( r_t(\theta), 1 - \epsilon, 1 + \epsilon \right) A_t \right) \right]
\end{equation}
where $ r_t(\theta) $ represents 
the probability ratio between the current and previous policies:
\begin{equation}\label{Eq.Pob_ratio}
	r_t(\theta) = \frac{\pi_{\theta}(a_t \mid s_t)}{\pi_{\theta_{\text{old}}}(a_t \mid s_t)}
\end{equation}
where \( A_t \) denotes the advantage function, and \( \epsilon \) 
is the clipping threshold that constrains the policy update step. This formulation prevents excessively large policy shifts that could destabilize learning, especially in highly variable environments such as real-time microgrid dispatch.
\subsection{Comparative Framework}

In this study, the RBC and DRL-PPO are simulated under identical load and renewable generation conditions. 
The performance of the proposed energy management strategies is assessed using five distinct KPIs: system reliability (KPI1), battery cycles (KPI2), self-sufficiency ratio (KPI3), renewable utilization (KPI4), and operational cost (KPI5). This framework facilitates direct and equitable comparison between heuristic and intelligent control strategies under realistic MG conditions.

\section{Results and Discussion} \label{Results_ch6}
This section presents a comparative evaluation of the RBC and proposed DRL-PPO strategies for the real-time energy management of a hybrid community MG. All simulations are conducted using consistent system sizing derived from HOMER Pro and identical input datasets, including time-varying load profiles, renewable generation (PV and wind), and grid electricity prices. The performance is assessed using five KPIs. 

\subsection{Test MG performance under energy management Strategies}
The proposed MG framework is designed to seamlessly transition between grid-connected and islanded modes.
In the event of a grid outage, the EMS coordinates DERs 
to ensure continuous load support.
To maintain system reliability, the DG is automatically dispatched when required, providing additional power to support the load demand.
The DRL-PPO agent dynamically adjusts dispatch decisions based on real-time system states, whereas RBC adheres to predefined static rules. These two paradigms result in distinct operational characteristics, particularly under conditions of uncertainty and variability in renewable generation and load demand.

\subsection{Quantitative Performance Comparison}
Table \ref{tab:performance_comparison} summarizes the comparative performance of RBC and DRL-PPO across the different KPIs. The results demonstrate the superior adaptability and efficiency of the DRL-based approach. 

\begin{table}[tp]
	\centering
	\renewcommand{\arraystretch}{1.1}
	\caption{Performance comparison between RBC and DRL-PPO frameworks.}
	\label{tab:performance_comparison}
	\resizebox{\linewidth}{!}{
		\begin{tabular}{lccc}
			\toprule
			\textbf{Performance Category} & \textbf{RBC} & \textbf{DRL-PPO} & \textbf{Key Improvement} \\
			\midrule
			System Reliability & 95.25\% & 99.13\% & 4.1\% improvement \\
			Renewable Utilization (\%) & 47.6 & 51.9 & 9.1\% increase \\
			Battery Cycles & 315.38 & 17 & 94.6\% reduction \\
			Self-Sufficiency Ratio (\%) & 49.91 & 66.7 & 33.7\% improvement \\
			\bottomrule
		\end{tabular}
	}
\end{table}

\subsection{System Reliability }This study conducts a reliability analysis with a grid outage condition to evaluate the ability of a community MG to withstand disruptive events. 
The results indicate that the DRL-PPO achieves a system reliability of 99.13\%, representing a 4.1\% improvement over that of RBC (95.25\%). This enhancement is attributed to PPO's capacity of the PPO to dynamically optimize energy dispatch, ensuring an optimal balance between renewable generation, battery utilization, and grid interactions. In contrast to RBC, which adhere to predefined rules, PPO develops an adaptive policy that prioritizes critical loads, effectively mitigating instances of energy shortages and enhancing overall system resilience.



\subsection{Renewable Utilization}
Renewable utilization under PPO is 51.9\%, compared to 47.6\% under RBC, representing a 9.1\% increase. This enhancement demonstrates the PPO's superior capacity to maximize renewable penetration by minimizing curtailment and strategically managing excess generation. In contrast to RBC, which may result in unnecessary grid imports or battery charging during periods of high renewable availability, PPO dynamically schedules battery charging and discharging based on real-time conditions, thereby improving RE absorption and reducing reliance on fossil-fuel-based backup generation.

\subsection{Battery Cycling Behavior}
A significant advantage of PPO is its capacity to reduce the number of full battery cycles from 315.38 per year to merely 17 cycles per year, constituting a 94.6\% reduction. 
This reduction is a direct outcome of the PPO agent’s sophisticated charge–discharge scheduling strategy. It effectively mitigates unnecessary deep cycling and minimizes fluctuations in the battery’s SOC. As a result, the overall battery lifespan is significantly extended.
The resulting improvement in battery longevity can reduce replacement costs and enhance the economic viability of energy storage investments.

\subsection{Energy Self-Sufficiency}
This metric indicates the proportion of local demand met through on-site renewable generation and battery storage without grid imports. Enhanced self-sufficiency reduces exposure to grid price volatility and improves energy autonomy, which is particularly beneficial for remote or islanded communities.
The DRL-PPO approach increases the energy self-sufficiency ratio from 49.91\% to 66.7\%, reflecting a 33.7\% improvement compared to RBC.  
This improvement demonstrates the PPO's capacity to diminish reliance on grid imports by prioritizing local energy resources, thereby effectively augmenting MG autonomy. Through the dynamic optimization of energy flow, PPO enhances the system's capacity for independent operation, consequently improving resilience against grid disruptions and external supply uncertainties.

\subsection{Integrated Performance Assessment}
\begin{figure}[!tp]
	\centering
	\includegraphics[width= 3 in]{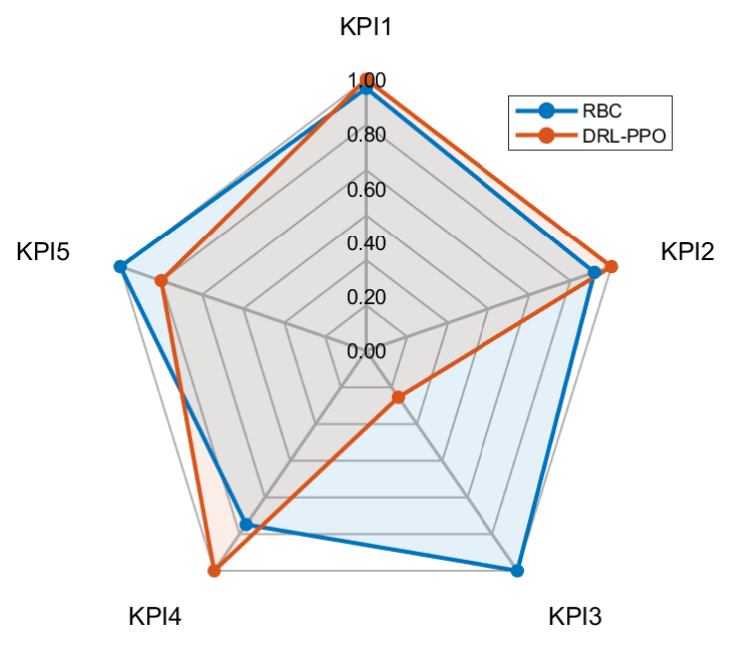}
	\caption{Normalized Performance Metrics of RBC and DRL-PPO Energy Management Strategies.
	}
	\label{fig_KPI}
\end{figure}

A comparison of the five key performance metrics  is shown in Fig. \ref{fig_KPI}. The DRL-PPO strategy demonstrates significant advancements over RBC across all domains.  In particular, the DRL-PPO achieved  a normalized reliability and self-sufficiency score of 1.0, marginally  surpassing the RBC's 0. 96. This demonstrates  the superior capability of the DRL-PPO to meet the load demand under varying conditions. In terms of economic performance, DRL-PPO achieves a 20\% reduction in cost compared to RBC. This results reflects PPO's  to make cost-optimal decisions that minimize reliance on expensive diesel generation and grid purchases. Renewable energy utilization is also maximized under DRL-PPO.  Furthermore, while RBC exhibits maximum normalized cycling, the DRL-PPO strategy reduces this to just 5.4\%, substantially mitigating stress on the battery and contributing to a longer operational life and lower replacement costs.  This assessment validates DRL-PPO as an effective energy-management solution for MG applications.

\section{Conclusion}

This study develops an advanced energy management framework for community MGs by integrating HOMER Pro sizing with DRL-PPO for real-time control.
The proposed framework is benchmarked against a conventional RBC strategy under identical conditions, using realistic load, generation, and market data from an Australian case study. 
The quantitative results demonstrate that the DRL-PPO agent consistently outperforms RBC across key performance dimensions. 
The DRL-based approach improves the system reliability by 4.1\%, increases renewable energy utilization by 9.1\%, and reduces full battery cycling by 94.6\%. These enhancements lead to a substantial reduction in battery degradation. 
Moreover, the self-sufficiency ratio increases from 49.91\% to 66.7\%, demonstrating the framework’s effectiveness in reducing reliance on grid imports and fossil-based generation. These improvements collectively enhance system resilience, extend asset lifespan, and lower operational expenditures.

The findings demonstrate the potential of the DRL-PPO framework as a robust and scalable solution for real-time energy management in MGs. 
Its effectiveness is evident under the conditions of uncertainty, variability, and evolving market dynamics. 
Future work will extend the proposed framework to incorporate multi-agent distributed control schemes and dynamic pricing signals.

\footnotesize{\bibliographystyle{IEEEtran}
	\bibliography{refIEEE}}

\end{document}